# Analysis of the longitudinal space charge impedance of a round uniform beam inside parallel plates and rectangular chambers

L. Wang

*SLAC National Accelerator Laboratory, Menlo Park, CA 94025, USA*

Y. Li

*Department of Physics, Michigan State University, East Lansing, MI 48824, USA*

This paper analyzes the longitudinal space charge (LSC) impedances of a round uniform beam inside a rectangular and parallel plate chambers using the image charge method. This analysis is valid for arbitrary wavelengths and the calculation converges fast. The research shows that only a few of the image beams are needed to obtain a relative error less than 0.1%. The beam offset effect is also discussed in the analysis.

## I. INTRODUCTION

The LSC impedances are important factors that should be taken into account in the design and operation of modern particle accelerators, especially the ones with low energy and high beam intensities, which are prone to microwave instabilities induced by the LSC fields [1, 2]. It is also important for the Free-Electron Lasers (FELs) where the electron's energy is up to a few GeV but its peak current is on the order of kA [3]. Strong microbunching on the order of micrometers has been recently observed [4] in Linac Coherent Light Source (LCLS) [5] at SLAC National Accelerator Laboratory (SLAC). On the other hand, the LSC, as a positive factor, increases the current modulation and can be used as an amplifier for FEL [6]. Various LSC field models have been proposed and studied extensively by different methods in the existing literatures, such as Refs. [7-22]. The LSC impedances in the entire wavelength spectrum of a beam in free space [8] and inside a round chamber [13-16] have been intensively studied. The LSC impedance of a beam inside a round chamber in the long-wavelength limit is also well-known [17-21].

In many accelerators, the configurations of the real beam and chamber system can be simplified as a field model consisting of a round beam moving inside a rectangular chamber. When the aspect ratio of the cross-section of the rectangular chamber is large, it can be simplified further as a pair of parallel plates [24]. If the beam has longitudinal density modulations, due to mismatch of the cross-sectional shapes between the beam and chamber, the exact closed-form solutions to the three-dimensional (3D) space charge fields and the corresponding LSC impedances of the above two cases cannot be obtained by the method of separation of variables. Till now, few works were carried out for obtaining the LSC impedances in the entire wavelength spectrum of the field models in a rectangular chamber and between a pair of parallel plates. Though Ref. [7] used the method of conformal mapping and Faraday's law applied to a rectangular integral loop to derive the analytical LSC fields and impedances of a round beam in rectangular chamber, it neglected the 3D field effects when the density perturbation wavelength is small. Hence, this method and results are only valid in the long-wavelength limit and are not appropriate for the study of microwave or microbunching instabilities. Since in most accelerators the ratios of the transverse chamber dimensions to the beam diameters are large, the Ref. [12] provides the approximate solutions to the field models of a round beam with planar and rectangular boundary conditions, assuming the 3D image charge fields of the round beam can be approximated by the image fields of a line charge. The resulting LSC fields and impedances are valid for the whole perturbation wavelength spectrum. When the ratio of the beam diameter to the transverse chamber dimension approaches unity, the relative errors of the approximated LSC impedances will become larger. In addition, the LSC field of a round beam between parallel plates has been studied in Ref. [22] but the results are only valid in the long-wavelength limit.

This paper proposes an image charge method to calculate the LSC impedances of a round beam between parallel plates and inside a rectangular chamber. It is well-known that the solutions to the LSC fields of a round transversely uniform beam with sinusoidal line charge density modulations in free space are available in a closed form [8]. If the beam were placed inside a rectangular chamber or between parallel plates, due to planar symmetry and mirroring, the associated total image charge fields of the infinite chain (parallel plates model) or grid (rectangular chamber model) of the image beams can be calculated by simple summation. Adding the self-fields of the round beam in free space, the total LSC fields and impedances of the two models in discussion can be obtained. Through case study, we found the calculated LSC fields and impedances converge pretty fast with the number of image beams. Usually only a finite and small number of image beams are needed to obtain the LSC fields and impedances with relative errors less than 0.1%. The resolution of the calculated LSC fields and impedances depend on the number of image beams used in the calculation, rather than the ratio of the beam diameter to the transverse chamber dimension, as was the case in Ref. [12].

The image charge method is one of the popular methods in the study of the charged particle field, especially for solving the Poisson's equation (i.g., Ref. [23]). In some literatures, the round beam is approximated by a line charge [12, 20] in calculation of the image charge field for simplicity. In this paper, the exact image beams are included in the calculation of the full-spectrum LSC impedance.

This paper is organized as follows. Section II briefly introduces the wave equations describing the space charge fields of the charged beam. Section III briefly introduces the numerical calculation method for general geometries of the beam and beam pipe. Section IV provides a short review for the LSC fields and impedances of a round beam in free space and inside a round chamber, respectively. Sections V and VI calculate the LSC impedances of a round beam between parallel plates and inside a rectangular chamber using the image charge method, respectively.

## II. WAVE EQUATIONS

The wave equation describing the electric field $\boldsymbol{E}$ is

$$\nabla^2 \boldsymbol{E} - \frac{1}{c^2}\frac{\partial \boldsymbol{E}}{\partial t} = \frac{\nabla \rho}{\varepsilon_0} + \mu_0 \frac{\partial \boldsymbol{J}}{\partial t}, \tag{1}$$

where $\epsilon_0$=8.85×10$^{-12}$ F/m and $\mu_0$= 4$\pi$×10$^{-7}$ N/A$^2$ are the permittivity and permeability of free space, respectively; $c = \frac{1}{\sqrt{\varepsilon_0\mu_0}} = 3 \times 10^8$ m/s is the speed of light in free space; ρ and $\boldsymbol{J}$ are the charge and the current densities, respectively , they obey the following continuity equation:

$$\frac{\partial\rho}{\partial t} + \vec{\nabla}\cdot\boldsymbol{J} = 0. \tag{2}$$

Assuming that the beam is moving with a constant longitudinal speed $\boldsymbol{\upsilon} = \beta c\hat{\boldsymbol{z}}$ along the $z$-axis, where $\beta$ is the relativistic speed, $\hat{\boldsymbol{z}}$ is the unit vector of the longitudinal coordinate, then $\boldsymbol{J}$ can be expressed as $\boldsymbol{J} = J\hat{\boldsymbol{z}}$. For a perturbed beam, its volume charge density $\rho$ and current density $J$ consist of unperturbed (DC) components and perturbed higher order harmonic components. Since the physics of the unperturbed components is trivial and does not contribute to the LSC field, we only need to focus on the physics associated with those higher order harmonic components. We will study one particular harmonic component with a frequency $\omega$ (or wave number $k$) in the rest of this paper and omit the subscript $k$ in the variables of fields, density and current for simplicity. The single harmonic component of the charge density, current density and beam current can be expressed using the wave assumption:

$$\rho(x,y,z,t) = \lambda_z\rho_\perp(x,y)e^{-i\omega(t-z/\upsilon)}, \tag{3a}$$

$$\boldsymbol{J} = \rho\boldsymbol{\upsilon} = \hat{\boldsymbol{z}}\beta c\lambda_z\rho_\perp(x,y)e^{-i\omega(t-z/\upsilon)}, \tag{3b}$$

$$I = \lambda_z\upsilon e^{-i\omega(t-z/\upsilon)} = \bar{I}e^{-i\omega(t-z/\upsilon)}, \tag{3c}$$

where $\lambda_z$ is the magnitude of the harmonic line charge density, $\rho_\perp$ is the transverse beam distribution function normalized by $\int\rho_\perp(x,y)dxdy = 1$. We will work with only the $z$-components of the vectors $\boldsymbol{E}$, $\boldsymbol{J}$ and ∇. Hence, the differentials of the longitudinal harmonic components of $J$ and $\rho$ can be expressed as

$$\mu_0\frac{\partial J_z}{\partial t} = -i\mu_0\omega\upsilon\lambda_z\rho_\perp e^{-i\omega\left(t-\frac{z}{\upsilon}\right)}, \tag{4}$$

$$\frac{1}{\varepsilon_0}\frac{\partial\rho}{\partial z} = \frac{i\omega\lambda_z}{\varepsilon_0\upsilon}\rho_\perp e^{-i\omega(t-z/\upsilon)}. \tag{5}$$

The $z$-component of the harmonic electric field can be written as

$$\mathbf{E}_z(x,y,z,t) = \hat{\mathbf{z}}\mathrm{E}_z(k,x,y)e^{-i\omega(t-z/\upsilon)}. \tag{6}$$

Substituting Eqs. (4-6) into Eq. (1), the amplitude of the longitudinal electric field $E_z(k, x, y)$ satisfies the following equation

$$\left(\nabla_\perp^2 - \frac{k^2}{\gamma^2}\right)\mathrm{E}_z = i\frac{k\lambda_z}{\varepsilon_0\gamma^2}\rho_\perp(x,y), \tag{7}$$

where $\nabla_\perp^2 = \frac{\partial^2}{\partial x^2} + \frac{\partial^2}{\partial y^2}$, $k = \frac{\omega}{\upsilon} = \frac{\omega}{\beta c}$ and $\frac{1}{\gamma^2} = 1 - \frac{\upsilon^2}{c^2} = 1 - \beta^2$.

The LSC impedance per unit length of an accelerator with circumference $L$ at an arbitrary transverse coordinate ($x$, $y$) is defined as

$$\frac{Z_{||}(k,x,y)}{L} = -\frac{E_z(k,x,y)}{\bar{\mathrm{I}}} = -\frac{E_z(k,x,y)}{\lambda_z\beta c}. \tag{8}$$

## III. FEM SIMULATION FOR ARBITRTARY GEOMETRY

Eq. (7) with arbitrary cross-sectional geometries of the beam and beam pipe can be solved numerically using the Finite Element Method (FEM) [25]. The FEM equation is

$$\left(\mathbf{M}+\frac{k^2}{\gamma^2}\mathbf{B}\right)\mathrm{E}_z=\mathbf{Q}\,, \tag{9}$$

with

$$M_{m,n}^e=\iint_{S^e}\left(\frac{\partial N_m}{\partial x}\frac{\partial N_n}{\partial x}+\frac{\partial N_m}{\partial y}\frac{\partial N_n}{\partial y}\right)dxdy\,, \tag{10}$$

$$B_{m,n}^e=\iint_{S^e}N_mN_n dxdy\,, \tag{11}$$

$$Q_m^e=-i\frac{kq_m}{\varepsilon_0\gamma^2}\,. \tag{12}$$

Here $\boldsymbol{M}$ is the stiffness matrix with matrix element $M_{m,n}^e$, $m$ and $n$ are the node indices of the finite element, $S^e$ is the integration boundary of the finite element. $N(x, y)$ is called the shape function (similar to the weighting factor) in FEM, by which the fields at a field point $P(x, y)$ within an element can be interpolated by the fields of its neighboring nodes. It is related to the coordinates of the field point $P(x, y)$ and the nodes of the element region. $q_m$ is the charge at the node $m$, which is proportional to $\lambda_z$. The current $I$ has the similar dependence on $\lambda_z$. Therefore, the LSC impedance given by Eq. (8) is independent of $\lambda_z$ as expected. The $E_z$ of Eq. (9) at all nodes satisfying equations Eqs. (9)-(12) and the boundary condition $E_z= 0$ on the chamber wall can be solved numerically. Then the corresponding longitudinal space charge impedances can be calculated using Eq. (8).

## IV. IN FREE SPACE AND INSIDE A ROUND CHAMBER

For an infinitely long round beam with uniform transverse density within beam radius $a$, its transverse density distribution function is

$$\rho_\perp(r)=\begin{cases}\frac{1}{\pi a^2} & (r\le a)\\ 0 & (r>a)\end{cases}, \tag{13}$$

where $r^2=x^2+y^2$. The general solution of Eq. (7) with the above beam distribution is [26]

$$E_Z(k,r)=\begin{cases}A_1I_0\left(\frac{kr}{\gamma}\right)+A_2K_0\left(\frac{kr}{\gamma}\right) & r>a,\\ A_3I_0\left(\frac{kr}{\gamma}\right)-i\frac{\lambda_z}{k\pi a^2\varepsilon_0} & r\le a.\end{cases} \tag{14}$$

where $I_0(x)$ and $K_0(x)$ are the 0$^{\text{th}}$ order modified Bessel functions of the first and second kinds, respectively. In free space, the field strength at $r\to\infty$ should be finite. Therefore $A_1$=0. The continuity conditions of the field and its derivative at $r=a$ give [8]

$$E_z^{free}(k,r)=\begin{cases}-i\frac{\lambda_z}{k\pi a^2\varepsilon_0}\left[1-\frac{ka}{\gamma}K_1\left(\frac{ka}{\gamma}\right)I_0\left(\frac{kr}{\gamma}\right)\right] & (r\le a),\\ -i\frac{\lambda_z}{k\pi a^2\varepsilon_0}\left[\frac{ka}{\gamma}K_0\left(\frac{kr}{\gamma}\right)I_1\left(\frac{ka}{\gamma}\right)\right] & (r>a).\end{cases} \tag{15}$$

where $I_1(x)$ and $K_1(x)$ are the 1$^{\text{st}}$ order modified Bessel functions of the first and second kinds, respectively. The superscript "*free*" on the left hand stands for "*free space*". With the property of $I_0(0)$=1, Eqs. (8) and (15) yield the LSC field and impedance per unit length on the beam axis ($r$=0) as

$$E_z^{free}(k,0)=-i\frac{\lambda_z}{k\pi a^2\varepsilon_0}[1-\frac{ka}{\gamma}K_1(\frac{ka}{\gamma})], \tag{16}$$

$$\frac{Z_{||}^{free}(k,0)}{L} = i\frac{Z_0}{k\pi a^2\beta}[1-\frac{ka}{\gamma}K_1(\frac{ka}{\gamma})]. \quad (17)$$

where $Z_0 = 1/\varepsilon_0 c \approx 377\ Ohms$ is the impedance of free space. Since the longitudinal electric field depends on the radial position, the LSC impedance also has the same dependence. The LSC impedance per unit length for arbitrary *r* within the beam (*r*<*a*) is given by

$$\frac{Z_{||}^{free}(k,r)}{L} = i\frac{Z_0}{k\pi a^2\beta}\left[1-\frac{ka}{\gamma}K_1\left(\frac{ka}{\gamma}\right)I_0\left(\frac{kr}{\gamma}\right)\right]. \quad (18)$$

Inside a round beam chamber with inner wall radius $r_w$, the continuity conditions at $r=a$ and the boundary condition on the chamber surface $E_z(r=r_w)=0$ determine the coefficients $A_1$, $A_2$ and $A_3$ in Eq. (14). Therefore the final solution of the longitudinal electric field is

$$E_z^{rd}(k,r) = \begin{cases} -i\frac{\lambda_z}{k\pi a^2\varepsilon_0}\left\{1-\frac{ka}{\gamma}\frac{I_0\left(\frac{kr}{\gamma}\right)}{I_0\left(\frac{kr_w}{\gamma}\right)}\left[K_1\left(\frac{ka}{\gamma}\right)I_0\left(\frac{kr_w}{\gamma}\right)+K_0\left(\frac{kr_w}{\gamma}\right)I_1\left(\frac{ka}{\gamma}\right)\right]\right\}, & (r\le a) \\ -i\frac{\lambda_z}{k\pi a^2\varepsilon_0}\frac{ka}{\gamma}I_1\left(\frac{ka}{\gamma}\right)\left[K_0\left(\frac{kr}{\gamma}\right)-\frac{K_0\left(\frac{kr_w}{\gamma}\right)}{I_0\left(\frac{kr_w}{\gamma}\right)}I_0\left(\frac{kr}{\gamma}\right)\right]. & (a<r\le r_w) \end{cases} \quad (19)$$

The superscript "*rd*" on the left hand side stands for "*round chamber*". The above equation gives the well-known LSC impedance per unit length of a round uniform beam inside a round beam chamber [13-16]

$$\frac{Z_{||}^{rd}(k,r)}{L} = i\frac{Z_0}{k\pi a^2\beta}\left\{1-\frac{ka}{\gamma}\frac{I_0\left(\frac{kr}{\gamma}\right)}{I_0\left(\frac{kr_w}{\gamma}\right)}\left[K_1\left(\frac{ka}{\gamma}\right)I_0\left(\frac{kr_w}{\gamma}\right)+K_0\left(\frac{kr_w}{\gamma}\right)I_1\left(\frac{ka}{\gamma}\right)\right]\right\}. \quad (20)$$

Using the identity of $< I_0(kr/\gamma) >= 2I_1(ka/\gamma)/(ka/\gamma)$, the *averaged* LSC impedance over the beam cross-section per unit length in free space and inside a round chamber can be derived easily from Eq. (18) and Eq. (20) as

$$\frac{\bar{Z}_{||}^{free}(k)}{L} = i\frac{Z_0}{k\pi a^2\beta}\left[1-2K_1\left(\frac{ka}{\gamma}\right)I_1\left(\frac{ka}{\gamma}\right)\right], \quad (21)$$

$$\frac{\bar{Z}_{||}^{rd}(k)}{L} = i\frac{Z_0}{k\pi a^2\beta}\left\{1-2\frac{I_1\left(\frac{ka}{\gamma}\right)}{I_0\left(\frac{kr_w}{\gamma}\right)}\left[K_1\left(\frac{ka}{\gamma}\right)I_0\left(\frac{kr_w}{\gamma}\right)+K_0\left(\frac{kr_w}{\gamma}\right)I_1\left(\frac{ka}{\gamma}\right)\right]\right\}. \quad (22)$$

In the long-wavelength limit ( $\gamma/(ka) \gg 1$), the *on-axis* LSC impedance of a round beam in free space is

$$\frac{Z_{||}^{free,LW}(k,0)}{L} = i\frac{kZ_0}{2\pi\beta\gamma^2}\left[\frac{1}{2}-C-\ln\left(\frac{ka}{2\gamma}\right)\right], \quad (23)$$

where *C*=0.577216 is the Euler's constant, the superscript "LW" stands for the "long-wavelength limit". The LSC impedance of a round beam centered inside a round beam pipe in the long-wavelength limit is [17-21]

$$\frac{Z_{||}^{rd,LW}(k)}{L} = i\frac{Z_0 k}{2\pi\beta\gamma^2}\left[C_1+\ln\left(\frac{r_w}{a}\right)\right], \quad (24)$$

where $C_1$=1/2 and 1/4 for the *on-axis* and *average* impedance, respectively. Fig. 1 shows the comparisons of the LSC impedance of a round beam in free space [Eq. (18) and Eq. (21)] and inside a round chamber [Eq. (20) and Eq. (22)]. Both the *on-axis* and *average* impedance are plotted for the purpose of comparison. The shielding of the beam chamber becomes more effective when $\gamma/(ka) > 1$. The formula of LSC impedance in the long-wavelength limit [(Eq. (24)] works well only when $\gamma/(ka) \gg 1$. Therefore, derivation of a full-spectrum analytical LSC impedance formula becomes necessary and important, which is one of the motivations of this paper.

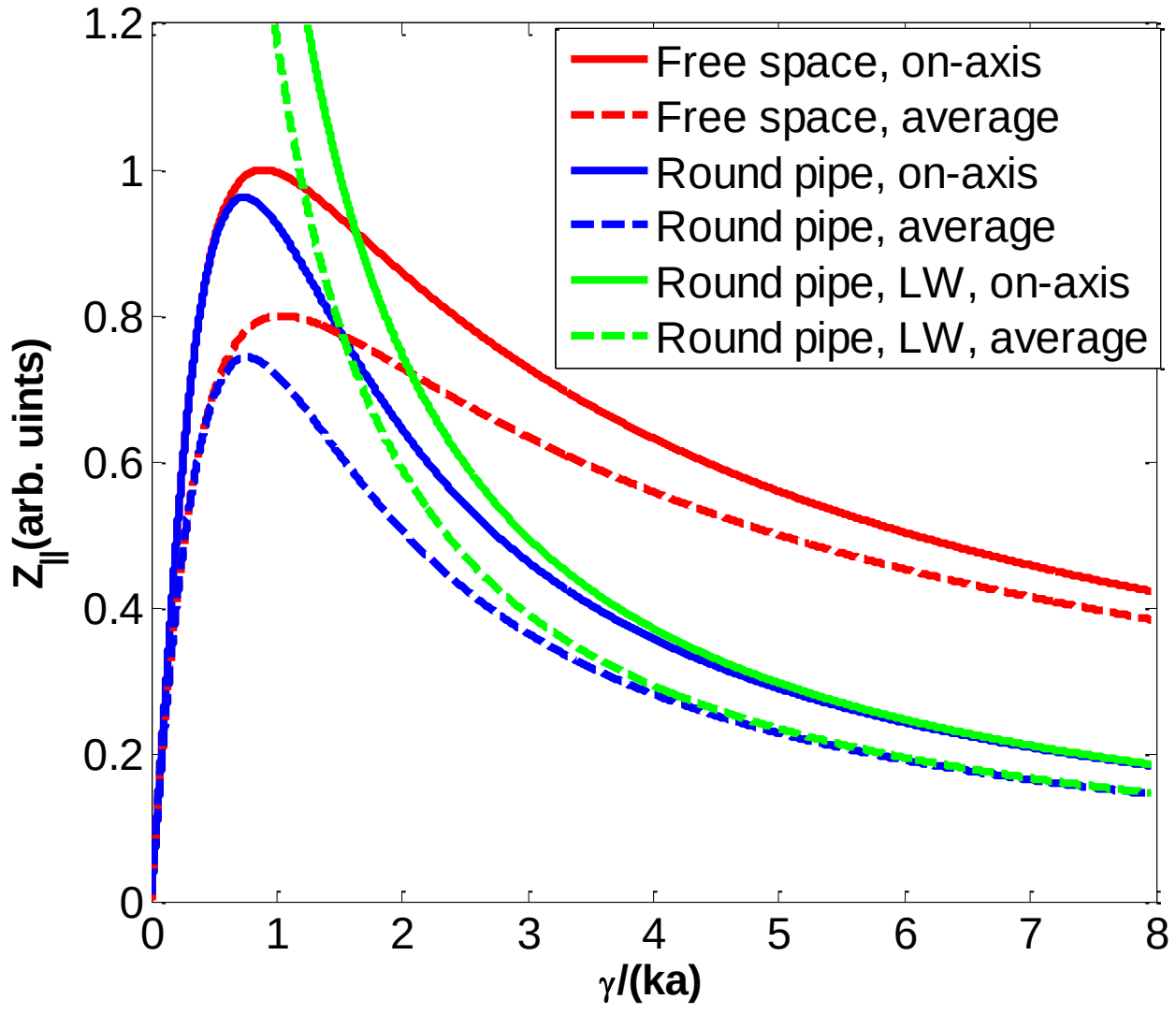

FIG. 1. Comparisons of the *on-axis* (solid lines) and *average* (dashed lines) LSC impedances of a round beam in free space and inside a round chamber. In the legend, 'LW' stands for the *long-wavelength* limit. The ratio of $r_w/a$ is 2 for this example.

## V. BETWEEN PARALLEL PLATES

Fig. 2 shows the cross-sections of a source beam between a pair of parallel plates and its images. We apply the image method by disregarding the boundary at each plate in a stepwise fashion to create the image map. Assuming the source beam has a line charge density of $\Lambda(z)=\lambda_z\cos(kz)$, the red and blue dots represent the beams with line charge densities of $\Lambda(z)$ and $-\Lambda(z)$, respectively. The distances of the source beam axis to the upper and lower plates are $d$ and $g$, respectively. The full separation between the two plates is $h=g+d$. If the source beam axis is located on the median plane between the two plates, then we have $g=d=h/2$.

Let us assume that the median plane of the parallel plate has vertical coordinate $y$=0 and a source beam is located at (0, $y_c$), as shown in Fig. 2. The image beams have vertical coordinates of $y_{img}(n) = nh + (-1)^n y_c$ and the corresponding line charge densities amplitudes of $\lambda_{z,imag}(n) = (-1)^n \lambda_z, n = 0, \pm1, \pm2 \cdots$. The images with the indices of $n$>0 and $n$<0 represent the ones above and below the parallel plates, respectively. The term with index of $n$=0 corresponds to the original source beam. For a round beam with uniform transverse density, its image beams have the same density distribution as the source beam as shown in Fig. 3. Note that the image beams within an arbitrary chamber may not always have the same shape as the source beam, for instance, the image beams within an elliptical chamber.

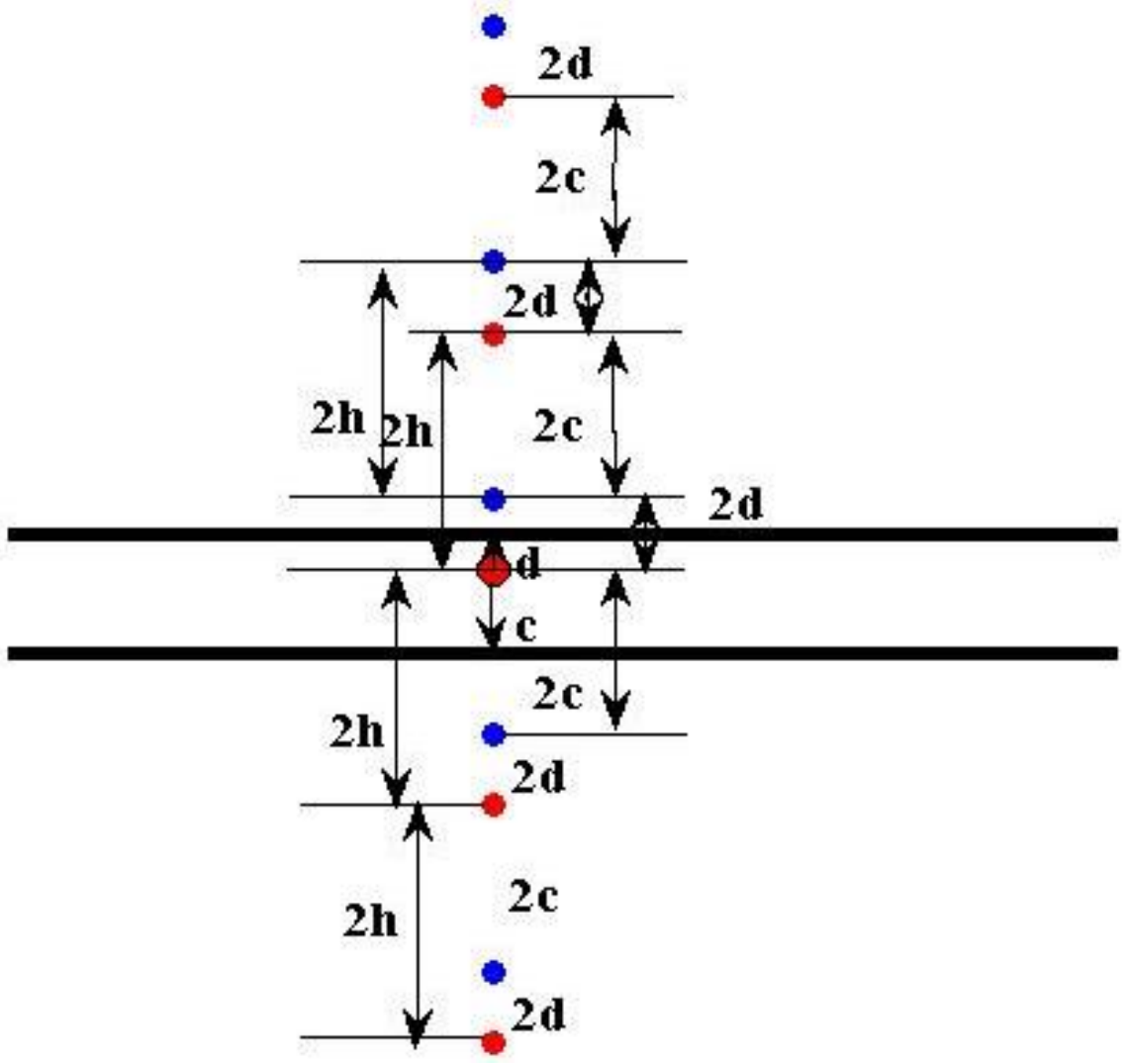

FIG. 2. Sequence of images of a source beam with vertical offset between a pair of parallel plates.

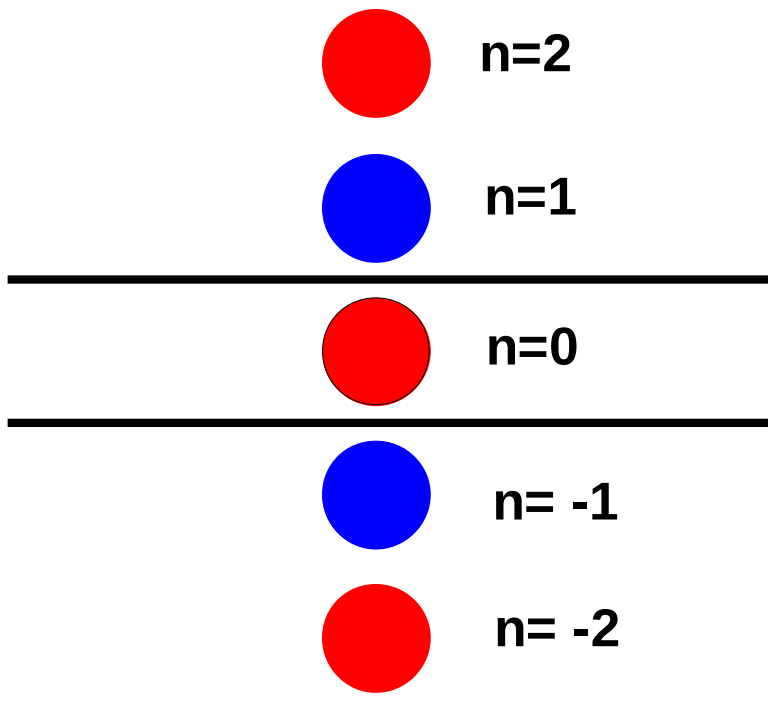

FIG. 3. Sequence of images of an infinitely long and uniform round beam centered between a pair of parallel plates. The image beams ($n$≠0) have the same shape and dimension as the source beam ($n$=0).

The field at any field point is equal to the sum of all image fields plus the self-field of the original source beam in free space. For instance, the LSC field at position ($x$, $y$) within the source beam is

$$E_z^{pp}(k,x,y) = \sum_{n=-\infty}^{\infty}(-1)^n E_z^{free}(k,R_n), \tag{25}$$

where $R_n = \sqrt{x^2 + (y_{img}(n) - y)^2}$. The superscript "*pp*" on the left hand of Eq. (25) stands for "*parallel plates*". The term with $n$=0 on the right hand side of Eq. (25) is contributed from the original source beam. Using Eqs. (8), (15) and (25), we obtain the LSC impedance per unit length at position ($x$, $y$) inside the beam as

$$\frac{Z_{||}^{pp}(k,x,y)}{L} = i\frac{Z_0}{k\pi a^2\beta}\left[1 - \frac{ka}{\gamma}K_1\left(\frac{ka}{\gamma}\right)I_0\left(\frac{kR_0}{\gamma}\right) + \frac{ka}{\gamma}I_1\left(\frac{ka}{\gamma}\right)\sum_{\substack{n=-\infty \\ n\neq 0}}^{\infty}(-1)^n K_0\left(\frac{kR_n}{\gamma}\right)\right]. \tag{26}$$

Different from the free space case, the LSC field of a round beam between parallel plates is not axisymmetric due to the boundary shape. Therefore, the impedance is a function of ($x$, $y$) instead of $r$ exclusively.

Fig. 4 shows an example of the calculated *on-axis* impedance with different number of image beams. The full height of the parallel plate is $h$=4$a$ in this example. The image beams affect the impedance mostly in the long-wavelength regime where $\gamma/(ka) > 1$. This clearly explains why the shielding is effective only for the long wavelength perturbations. The impedance with $|n|$=1 has the smallest value because it includes only the first pair of image beams whose line charge densities are opposite to that of the original beam. On the other hand, the impedance with $|n|$=2 has the maximum value. Similarly, $|n|$=3 gives the second smallest one and $|n|$=4 gives the second largest one and so on. When $|n|$ is large enough the variation of the impedance is negligible, because the image beams are far away from the source beam and the field finally converges.

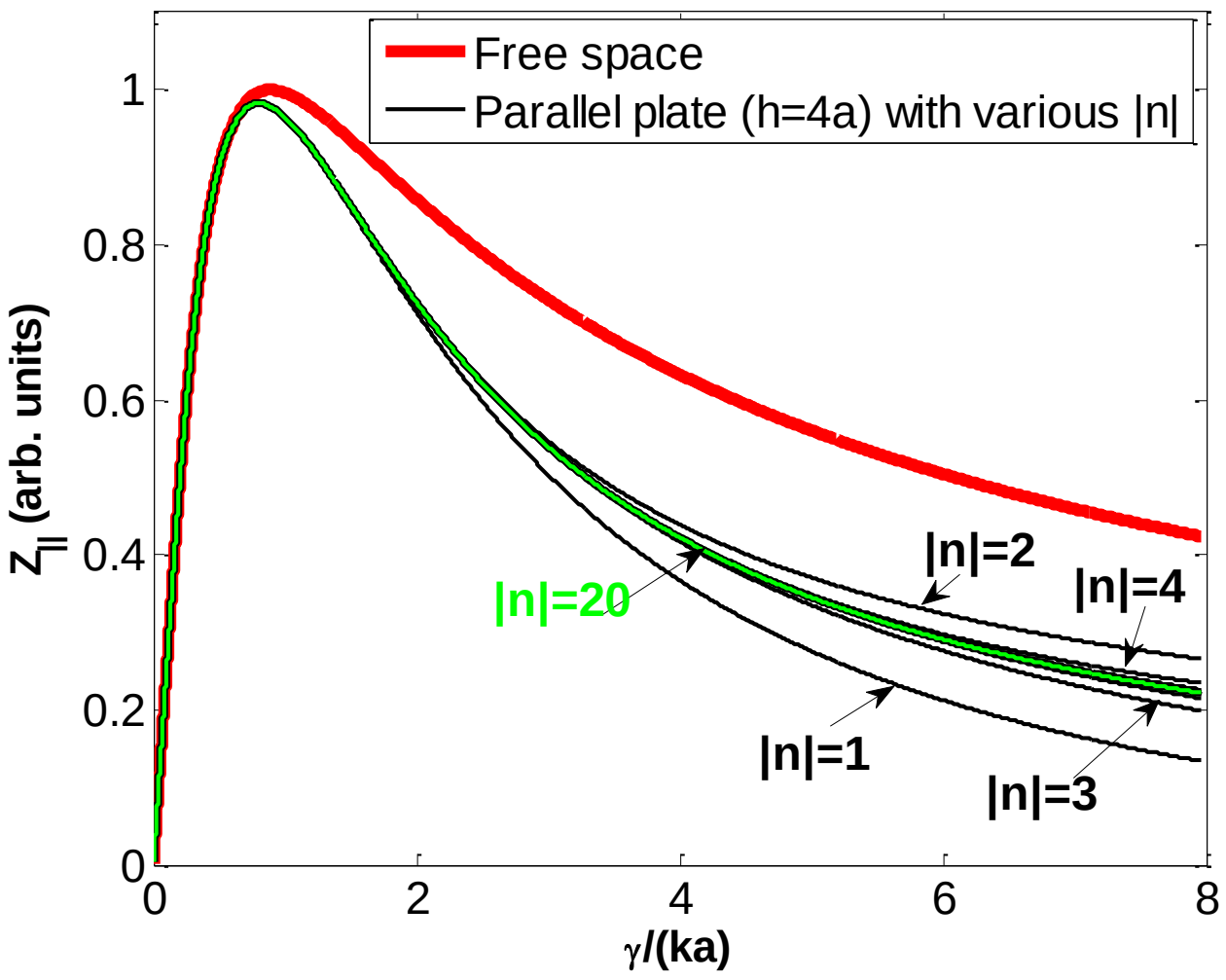

FIG.4. The effects of image bunches on the *on-axis* LSC impedance of a round transversely uniform beam centered between a parallel-plate chamber. The bunch has a radius of $a$=5 *mm*. The full height of the parallel plate is $h$=4$a$=20 *mm*. The family of black lines shows the calculated impedances with different maximum indices $|n|$ of image beams in Eq. (26) ranging from 1 to 20. The green line in the plot shows the calculated impedance with $|n|$=20.

Fig. 5 shows the variation of the *on-axis* LSC impedance at $\gamma/(ka) = 8$ with the maximum indices |n| of image beams ($h/a$=4). The calculated impedance (blue curve) oscillates at small $|n|$ and converges to one value at large $|n|$. The relative errors (red curve) in the plot are the absolute values of the relative errors. The true errors oscillate with $|n|$ and have negative sign for odd numbers of $|n|$. The absolute magnitudes of the relative errors decay exponentially which implies that the field converges with $|n|$ quickly. The case of $|n|$=15 gives a relative error on the order of magnitude of $10^{-4}$.

When the beam axis has vertical offsets with respect to the chamber median plane (see Fig. 2), the *on-axis* LSC impedance at long wavelength is slightly reduced as shown in Fig. 6 assuming $h$=10$a$ and the beam offsets range from *a to* 4*a*. The beam offset only affects the long-wavelength impedance for the same mechanism as the shielding effect. When the beam axis shifts vertically from the chamber median plane, one group of images with line charge densities of $-\Lambda(z)$ (with $y_{img}$>0 in Fig. 2) moves closer to the beam while another group of images with line charge densities of $-\Lambda(z)$ (with $y_{img}$<0 in Fig. 2) moves away from it as shown in Fig. 2. While the distance of the image beams with line charge densities of $\Lambda(z)$ to the source beam doesn't change when the source beam is shifted. The net effect from the re-distribution of the image beams with line charge densities of $-\Lambda(z)$ is small due to the cancellations of all the image beams. In general case, the effect of beam offset is negligible if the offset amplitude is small compared to the aperture of beam pipe.

The LSC impedance of a round beam centered between parallel plates in the long-wavelength limit can be derived from Ref. [22] as

$$\frac{Z_{||}^{pp,\ LW}(k)}{L} = i\frac{Z_0 k}{2\pi\beta\gamma^2}\left[C_1 + ln\left(\frac{2h}{\pi a}\right)\right]. \tag{27}$$

where $C_1 = 1/2$ for the *on-axis* impedance and $C_1 = 1/4$ for the *average* one, respectively. Fig. 7 shows the comparisons of the *average* LSC impedances of a round beam with beam radius $a$ = 0.5 *cm* under different boundary conditions and in different wavelength limit. We used the 20 *keV* ($\beta \approx 0.0046$, $\gamma \approx 1.0$) coasting $H_2^+$ beam in the Small Isochronous Ring (SIR) at Michigan State University (MSU) [24] in this example. The circumference of SIR is 6.58 *m*. The analysis agrees well with the simulation in the entire wavelength spectrum; while the impedance formula of the long-wavelength limit works well only in the case of $\gamma/(ka) > 11.14$ (with the charge density perturbation wavelength $2\pi/k = 35cm$ in Fig. 7).

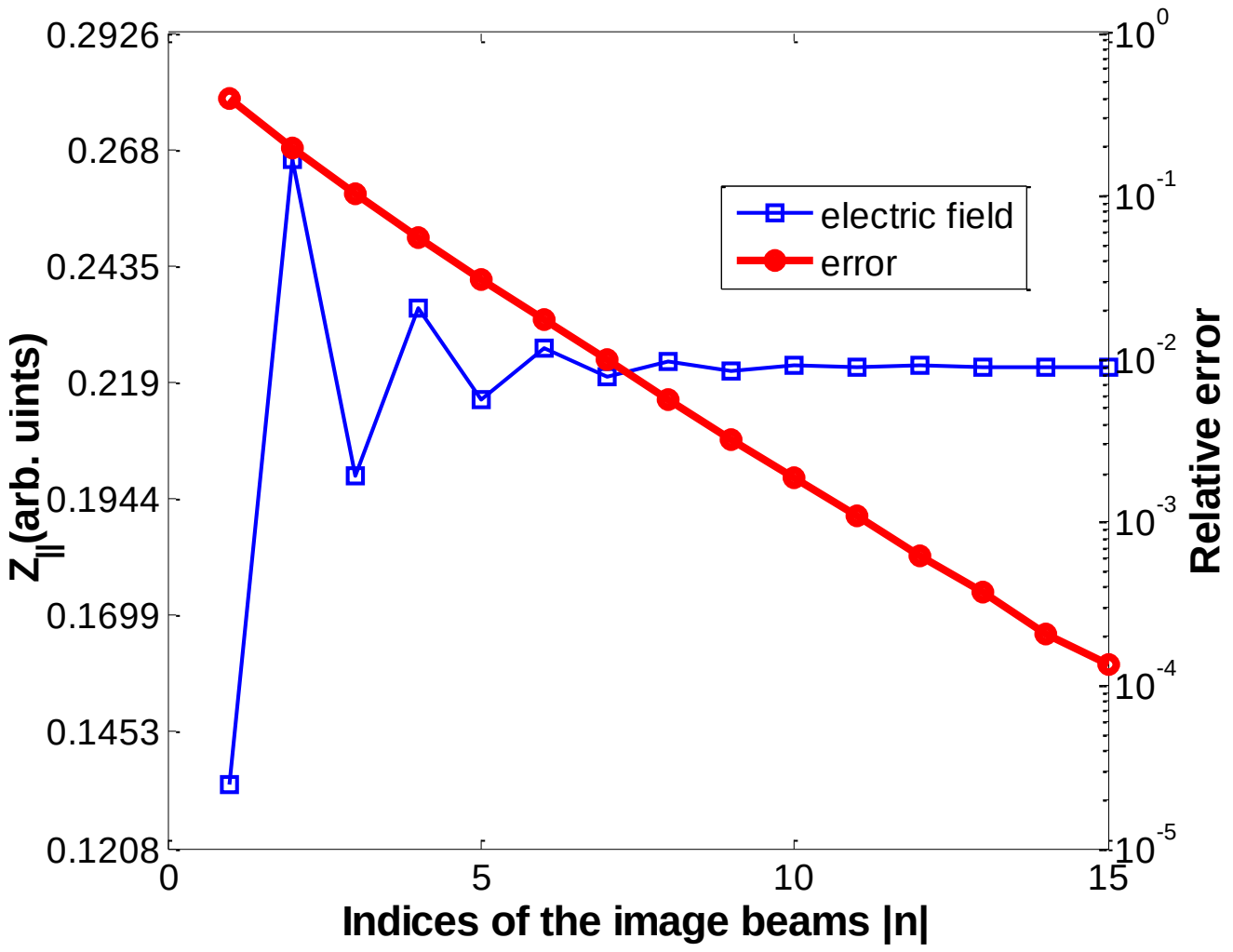

FIG. 5. Convergence of the *on-axis* LSC impedance with the maximum indices of image beams for a parallel-plate chamber. The bunch has a radius of *a*=5 *mm* and the full separation of the parallel plate is $h$=4$a$=20 *mm*. This result corresponds to the case $\gamma/(ka)$=8 as shown in Fig. 4 where the image charge effect is relatively larger.

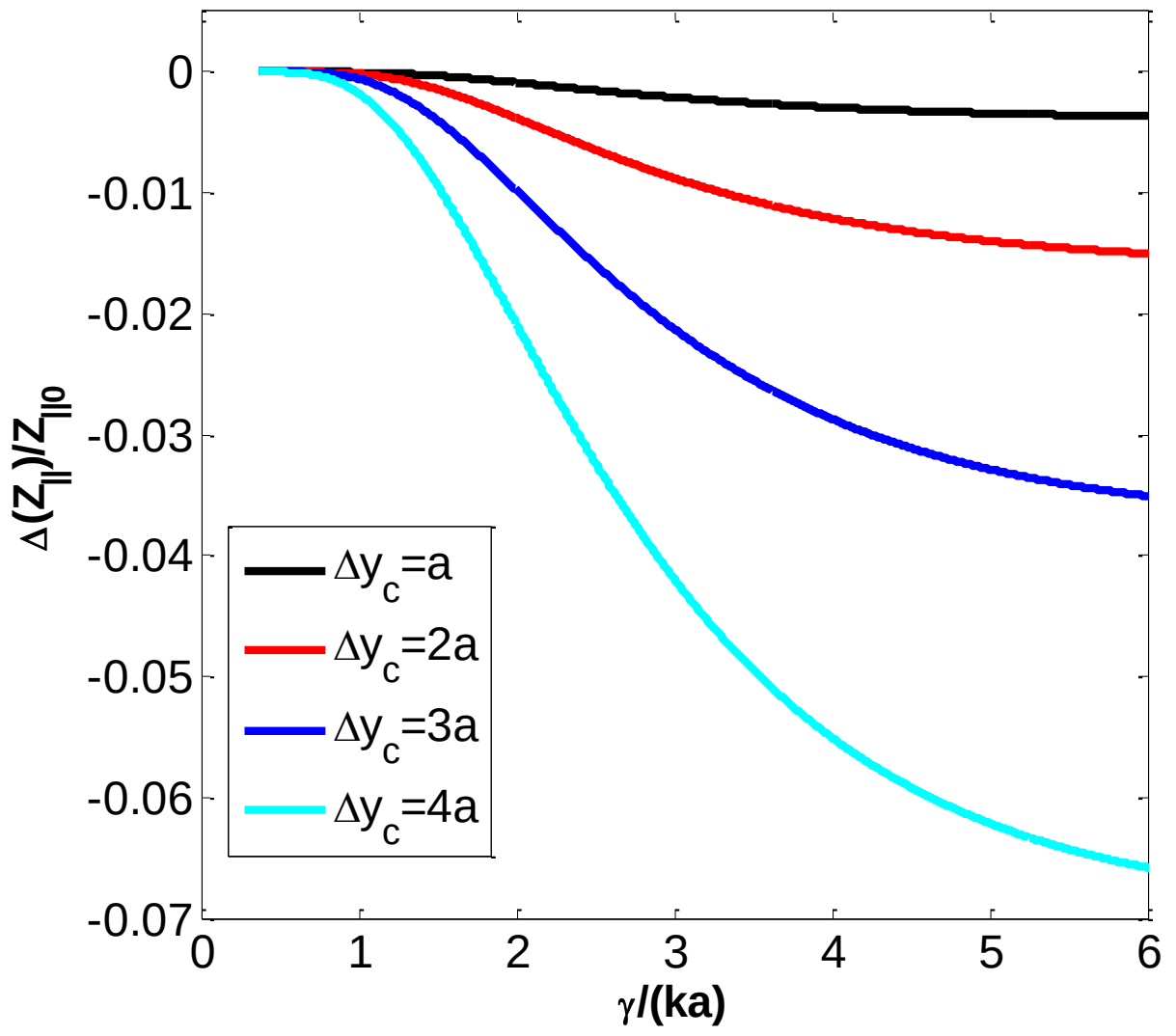

FIG. 6. Effects of vertical beam axis offset ($\Delta y_c$) on the *on-axis* LSC impedance of a parallel-plate chamber with $h$=10$a$. The relative impedance difference $\Delta Z_{\|}$ is normalized by the *on-axis* LSC impedance $Z_{\|0}$ with $\Delta y_c$=0. The impedance at long wavelength is slightly reduced when the beam axis has small vertical offsets.

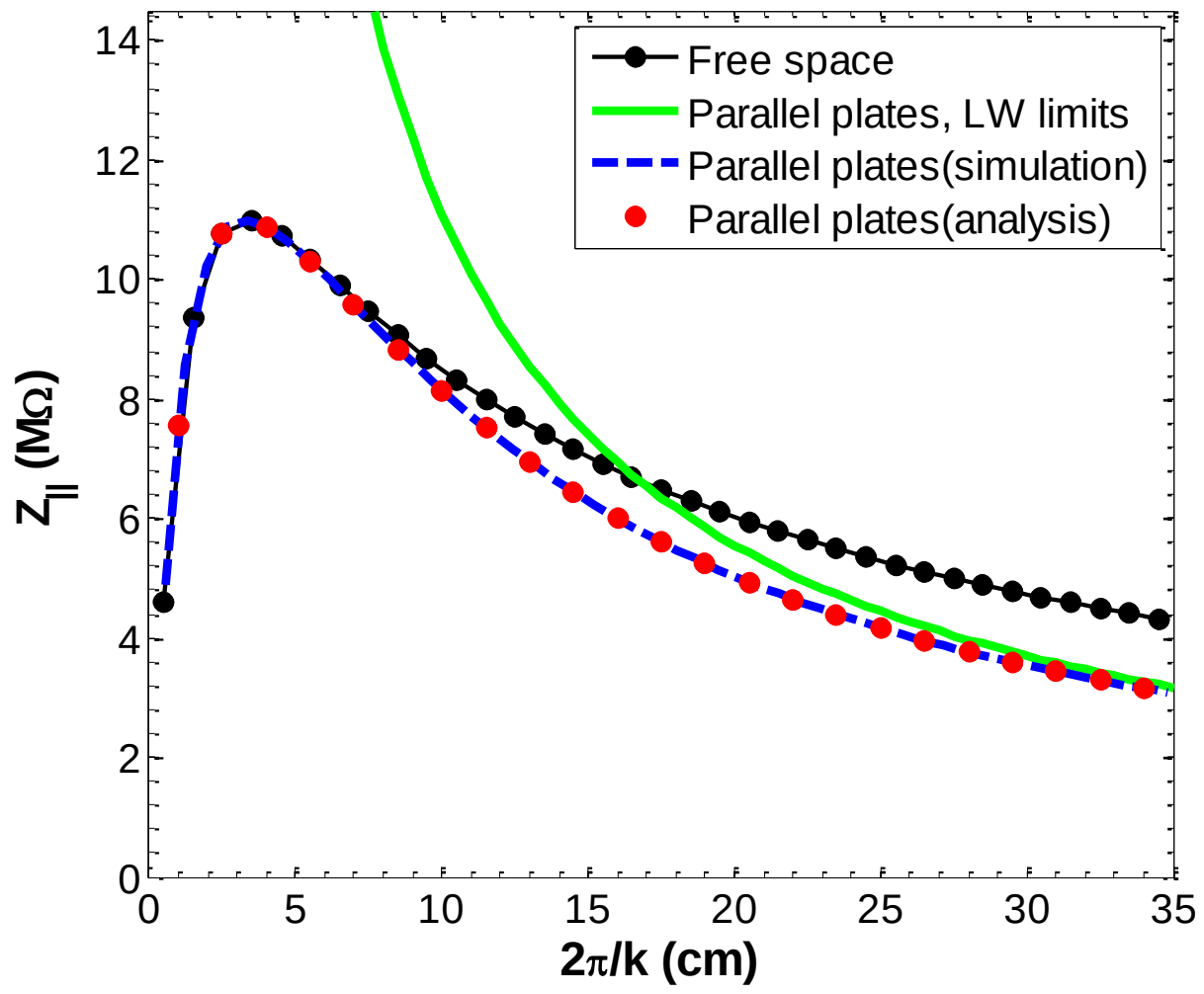

Fig. 7. Comparisons of the *average* LSC impedances of a *round* SIR beam with beam radius $a$ = 0.5 cm under different boundary conditions and in different wavelength limit. The full height of the parallel plates is 4.8 *cm*. In the legend, 'Free space', and 'Parallel plates' are boundary conditions; 'LW limits' stands for the *long-wavelength* limits; '(analysis)' and '(simulation)' stand for the theoretical result and simulation (FEM) method, respectively. The theoretical result is consistent well with the simulation in the entire wavelength spectrum.

## VI. INSIDE A RECTANGULAR CHAMBER

Consider an infinitely long, transversely uniform round beam inside a rectangular conducting structure with full width $w$ and full height $h$. We assume the axes of the rectangular chamber and the beam are located at (0, 0) and ($x_c$, $y_c$), respectively. Fig. 8 shows the image beams of a source beam centered inside a rectangular chamber indicated by the solid black rectangle. The red and blue dots represent the beams with line charge densities of $\Lambda$(z) and $-\Lambda$(z), respectively.

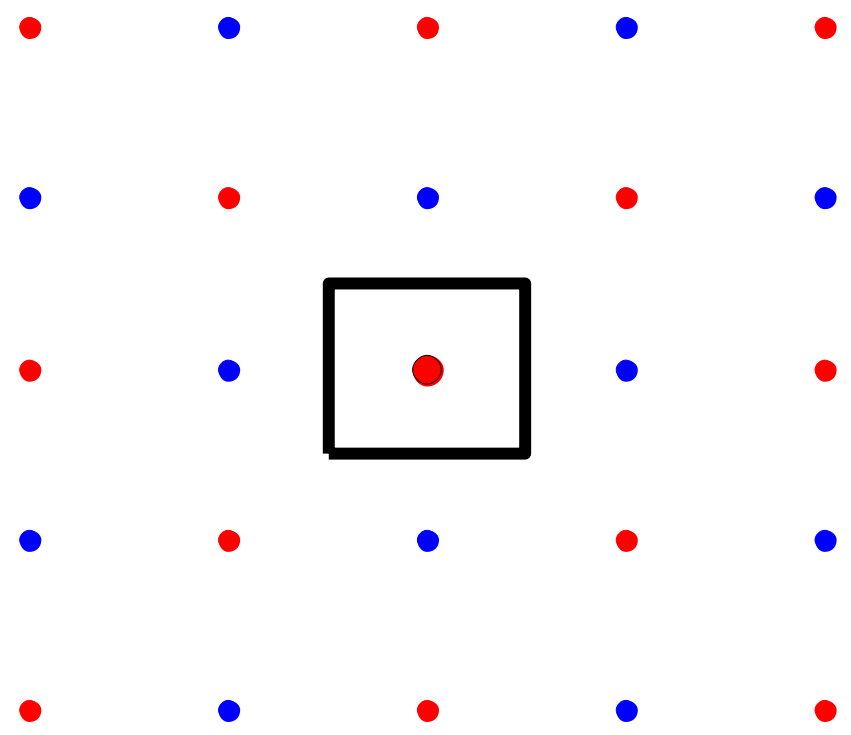

FIG. 8. 2D grid of images of a source beam centered inside a rectangular chamber.

Through planar symmetry and mirroring, we can determine the exact coordinates of images axes of a line beam inside a rectangular chamber as

$$x_{img}(m) = mw + (-1)^m x_c,$$

$$y_{img}(n) = nh + (-1)^n y_c,$$

$$\lambda_{z,imag}(m,n) = (-1)^{m+n}\lambda_z, m = n = 0, \pm 1,2,3\cdots \quad (28)$$

Here $m$ and $n$ represent the indices of the image grid points in horizontal and vertical directions, respectively. For instance, the indices $m$>0 ($m$<0) are for the images with $x$>0 ($x$<0). The image with the indices of $m = n = 0$ is just the original source beam. Similar to the case of parallel-plate chamber, the image beams of a round uniform beam still keep the same distributions and beam radius as those of the source beam, while their centers are given by Eq. (28). Therefore, the LSC field at ($x$, $y$) within the source beam inside a rectangular chamber is equal to the total LSC fields of the source beam and its image beams in free space

$$E_z^{rect}(k,x,y) = \sum_{m=-\infty}^{\infty}\sum_{n=-\infty}^{\infty}(-1)^{m+n}E_z^{free}(k,R_{m,n}), \quad (29)$$

with $R_{m,n} = \sqrt{\left(x_{img}(m,n) - x\right)^2 + \left(y_{img}(m,n) - y\right)^2}$. The superscript "*rect*" on the left hand of the equation stands for "*rectangular chamber*". Note that the original source beam effect is included with $m = n = 0$. Using Eqs. (8), (15) and (29), we obtain the LSC impedance per unit length at position ($x$, $y$) inside the beam as

$$\frac{Z_{||}^{rect}(k,x,y)}{L} = i\frac{Z_0}{k\pi a^2\beta}\left[1 - \frac{ka}{\gamma}K_1\left(\frac{ka}{\gamma}\right)I_0\left(\frac{kR_{0,0}}{\gamma}\right) + \frac{ka}{\gamma}I_1\left(\frac{ka}{\gamma}\right)\sum_{\substack{m=-\infty \\ m\neq 0}}^{m=\infty}\sum_{\substack{n=-\infty \\ n\neq 0}}^{n=\infty}(-1)^{m+n}K_0\left(\frac{kR_{m,n}}{\gamma}\right)\right]. \quad (30)$$

The LSC impedance per unit length of a round uniform beam centered inside a rectangular chamber in the long-wavelength limit is [7]

$$\frac{Z_{||}^{rect,\ LW}(k)}{L} \approx\ i\frac{Z_0 k}{2\pi\beta\gamma^2}\left[C_1 + ln\left(\frac{2h}{\pi a}\tanh\left(\frac{\pi w}{2h}\right)\right)\right],\ \frac{w}{h} > 1 \qquad (31)$$

where $C_1 = 1/2$ for the *on-axis* impedance and $C_1 = 1/4$ for the *average* one, respectively. The above formula is a good approximation for a rectangular chamber with *w*/*h*>1. When *w*/*h* is about 1, the exact but more complicated formula [7] should be used.

Fig. 9 shows the comparison of the *on-axis* LSC impedances obtained by the theoretical calculations and the simulations using a Finite Element Method (FEM) code. We used the 20 *keV* SIR beam in this example. The cross-section of the SIR chamber is rectangular with *w*=11.4 *cm* and *h*=4.8 *cm*. The maximum indices of the image beam used in the calculation are |*m*|=|*n*|=10. The *on-axis* LSC impedances of the beam with four different beam radii are compared to each other. The analytical results (solid lines) and FEM results (circles) perfectly overlap each other demonstrating excellent agreements in all cases in the entire wavelength spectrum. However, the long-wavelength-limit formula significantly overestimates the impedance at short wavelength as shown by the dashed lines in Fig. 9. Therefore, it is essential to use a more accurate full-spectrum LSC impedance formula in the study of beam instability.

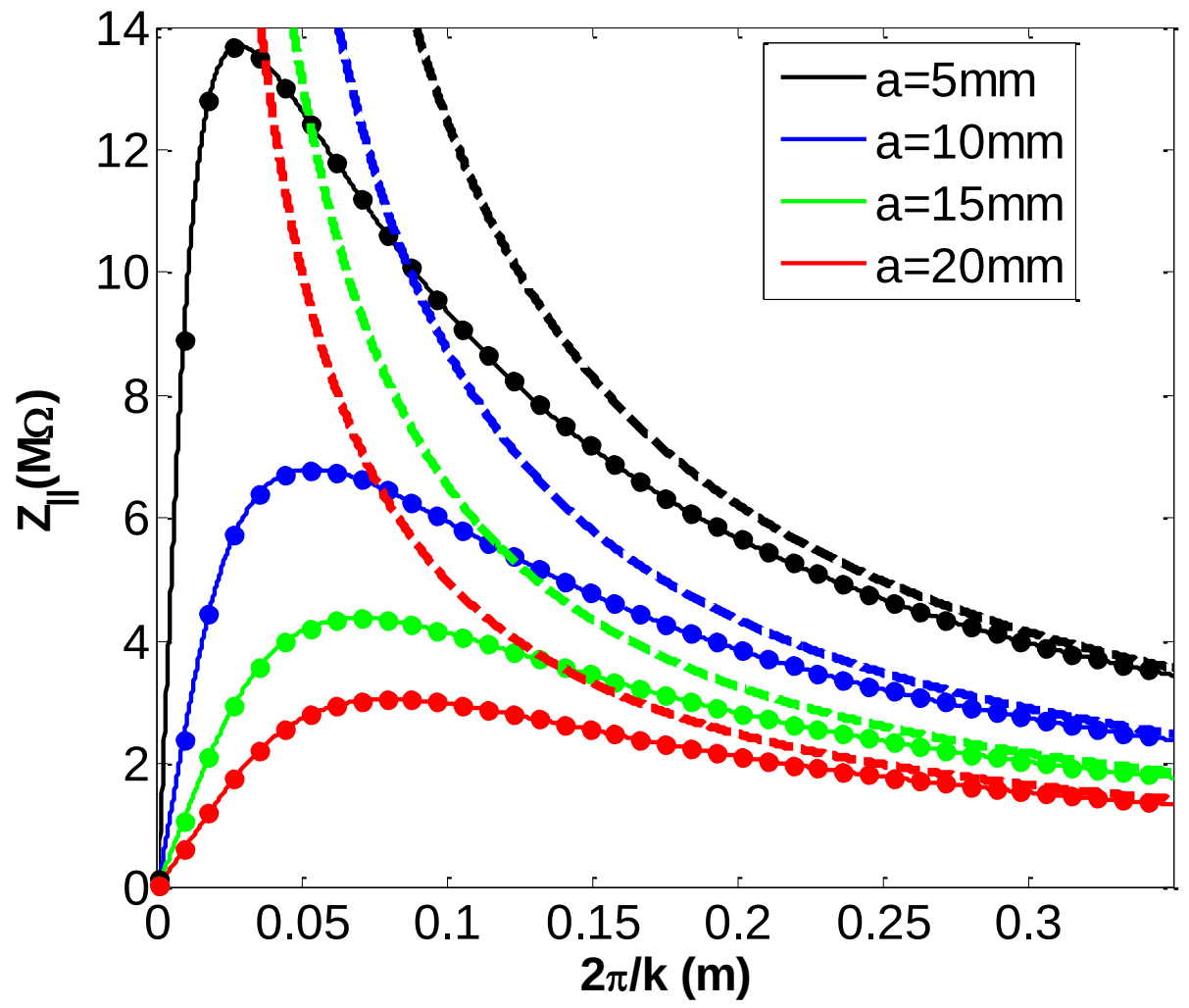

FIG.9. Comparisons of the *on-axis* LSC impedance of a round transversely uniform beam centered inside the SIR rectangular chamber between analysis (solid lines) and FEM simulations (solid dots). The LSC impedances predicted by the long-wavelength-limit formula (dashed lines) are also shown for comparisons. The parameters used in the calculations are *w* = 11.4 *cm*, *h* = 4.8 *cm*, the variable beam radii are *a* = 0.5, 1.0, 1.5 and 2.0 *cm* in this study. The circumference of SIR is 6.58 *m*.

Fig. 10 shows the shielding effect of a square chamber with *w*=*h*=4*a*, 6*a*, 8*a*,…16*a*. The shielding effect is negligible at short wavelengths with $\gamma/ka$<1.0, while it becomes more noticeable at long wavelengths.

For a fixed beam radius, the shielding effect becomes weaker as the chamber aperture is enlarged because the image beams are farther away from the source beam.

Fig. 11 shows the effect of beam offset on the *on-axis* impedance in a square chamber with $w = h = 10a$. The beam offset reduces the impedance and this reduction depends on the wavelength as shown in Fig. 11(a) where the vertical offset is zero. The beam offset effectively reduces the impedance at long wavelength regime due to the chamber shielding effect. Fig.11 (b) is for a particular wavelength with $\gamma/(ka) = 5$. The reduction of the LSC impedance is small when the beam offset is much smaller compared to the aperture of the beam pipe. However, the reduction of the impedance becomes more pronounced when the beam offset increases, because the shielding effect by the image beams roughly scales as $1/r$ ($r$ is the distance of the source beam axis to the surface of beam pipe). The impedance is about 50% smaller when the beam is close to the surface of the beam pipe as shown in the figure.

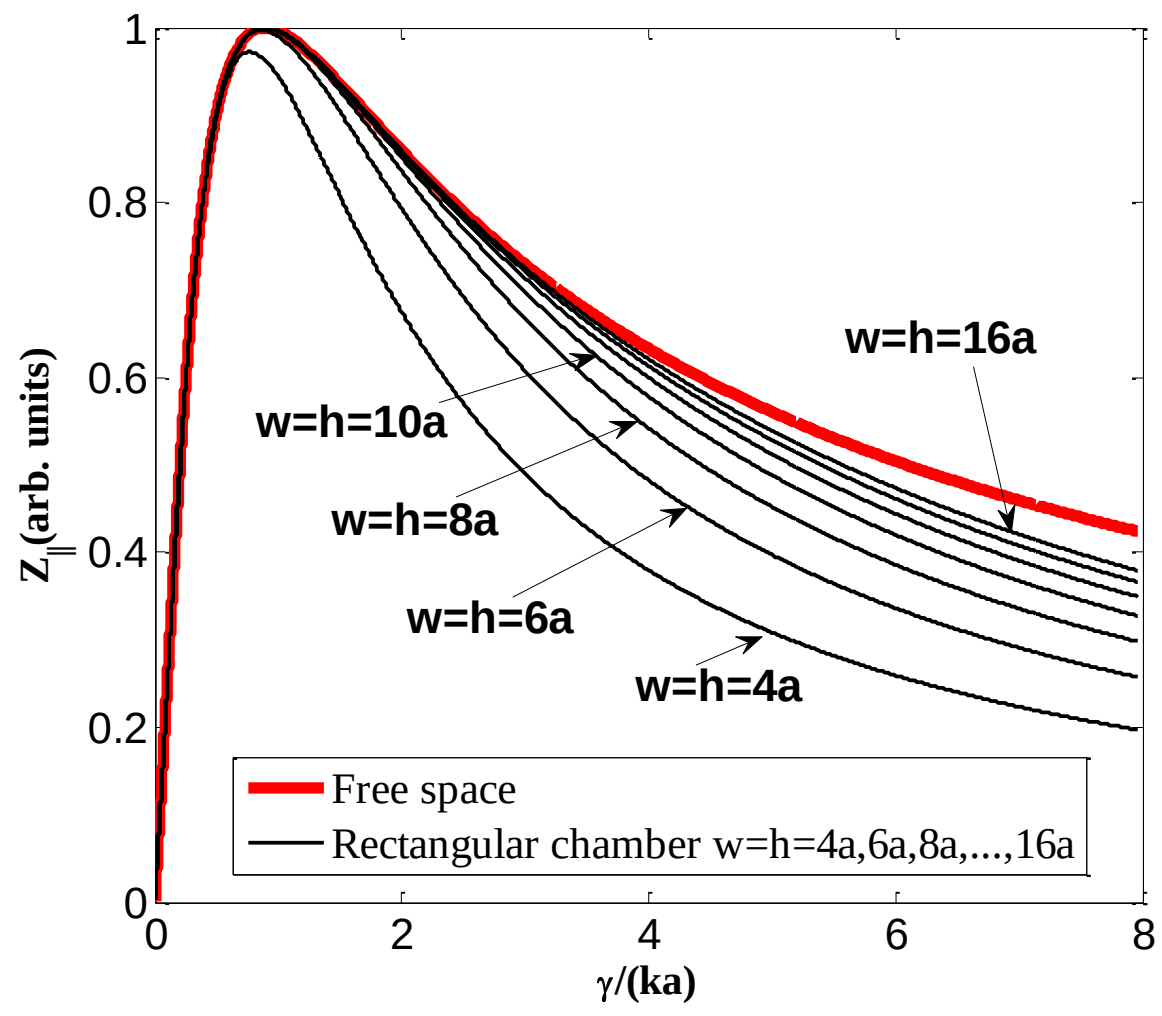

FIG. 10. The shielding effect of a rectangular chamber on the *on-axis* impedance. The round uniform beam has a radius of $a$, the full widths and heights of the chamber are $w$=$h$=4$a$, 6$a$, 8$a$, 10$a$, 12$a$, 14$a$ and 16$a$, which correspond to the lower to upper black lines as clearly shown on the right part of the plot.

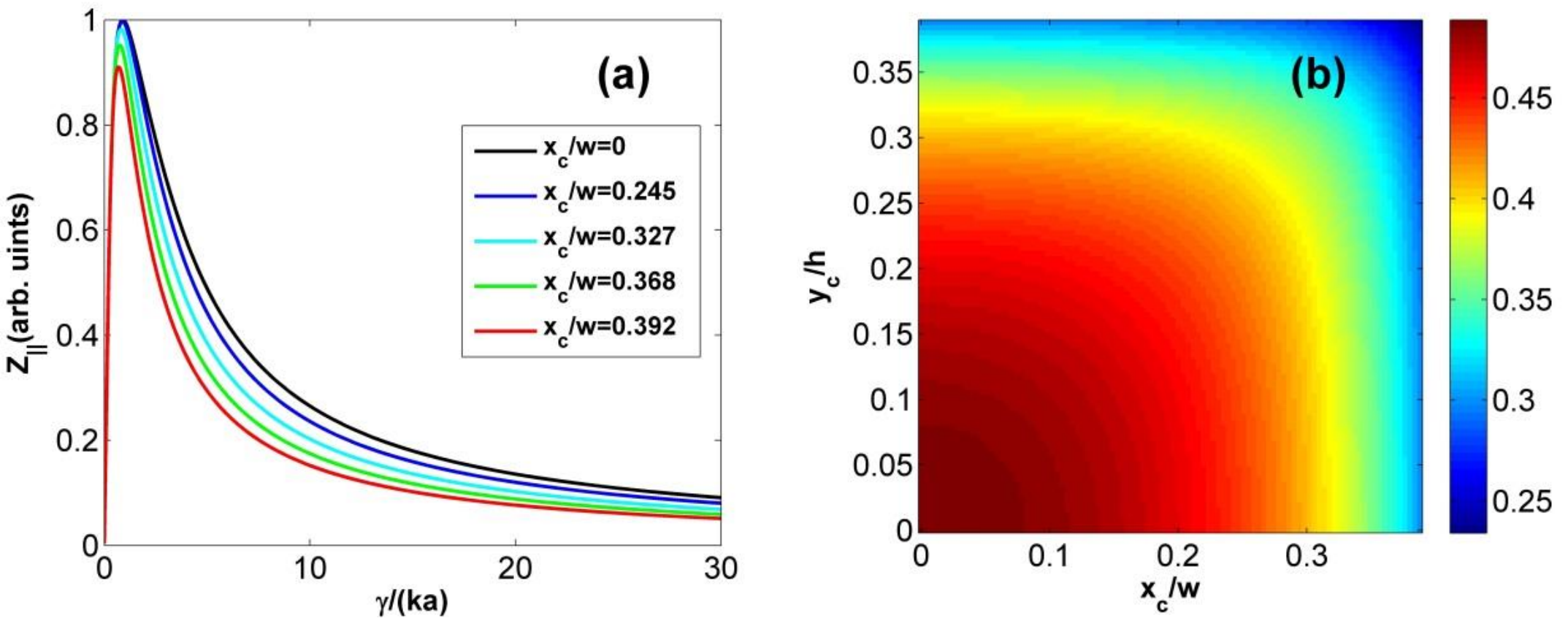

FIG. 11. The effect of beam offset ($x_c$, $y_c$) in a square chamber on the *on-axis* impedance. The impedance in the entire wavelength spectrum with different horizontal offsets $x_c$ ($y_c$=0) is shown in the plot (a); and the impedance at one particular wavelength $\gamma/(ka) = 5$ for various ($x_c$, $y_c$) is shown in the plot (b). The round uniform beam has a radius of $a$, the full width and height of the chamber are $w$=$h$=10$a$.

Fig. 12 compares the shielding effect on the *on-axis* impedance between a round chamber ($r_w = 2a$), parallel plates ($h$= $4a$) and rectangular chambers ($h$= $4a$, $w/h$=1and 2). The *on-axis* LSC impedance in free space and inside a round chamber is calculated using Eq. (17) and Eq. (20) for $r$=0, respectively. The round chamber has slightly stronger shielding than a square chamber resulting in a smaller *on-axis* LSC impedance. It is about 6% (1%) less than that of a round beam inside a square pipe when the pipe radius is 2$a$ (10$a$) at wavelength regime with $8 > \gamma/(ka) > 2$. When the aspect ratio of rectangular chamber $w/h$ is larger than 2, the shielding effect is very close to that of a parallel-plate model. The impedance of parallel-plate model at long wavelength regime is about 20% larger than that of the round-chamber model.

To compare the shielding effect in the long-wavelength limit we can define the geometry factor $g_0$ as

$$\frac{Z_{||}^{LW}(k)}{L} = i\frac{Z_0 k}{2\pi\beta\gamma^2} g_0 \ . \tag{32}$$

Comparing Eq. (24), Eq. (27) and Eq. (31) with the above equation, we can get the geometry factor $g_0$ of a transversely uniform round beam centered inside the round chamber, parallel plates and rectangular chamber as

$$g_0 = \begin{cases} C_1 + \ln\left(\frac{r_w}{a}\right), \ round\ chamber \\ C_1 + ln\left(\frac{2h}{\pi a}\right), \ parallel\ plates \\ C_1 + ln\left(\frac{2h}{\pi a}\tanh\left(\frac{\pi w}{2h}\right)\right), \ rect. chamber, \frac{w}{h} > 1 \end{cases} \ . \tag{33}$$

where $C_1 = 1/2$ for the *on-axis* impedance and $C_1 = 1/4$ for the *average* one, respectively.

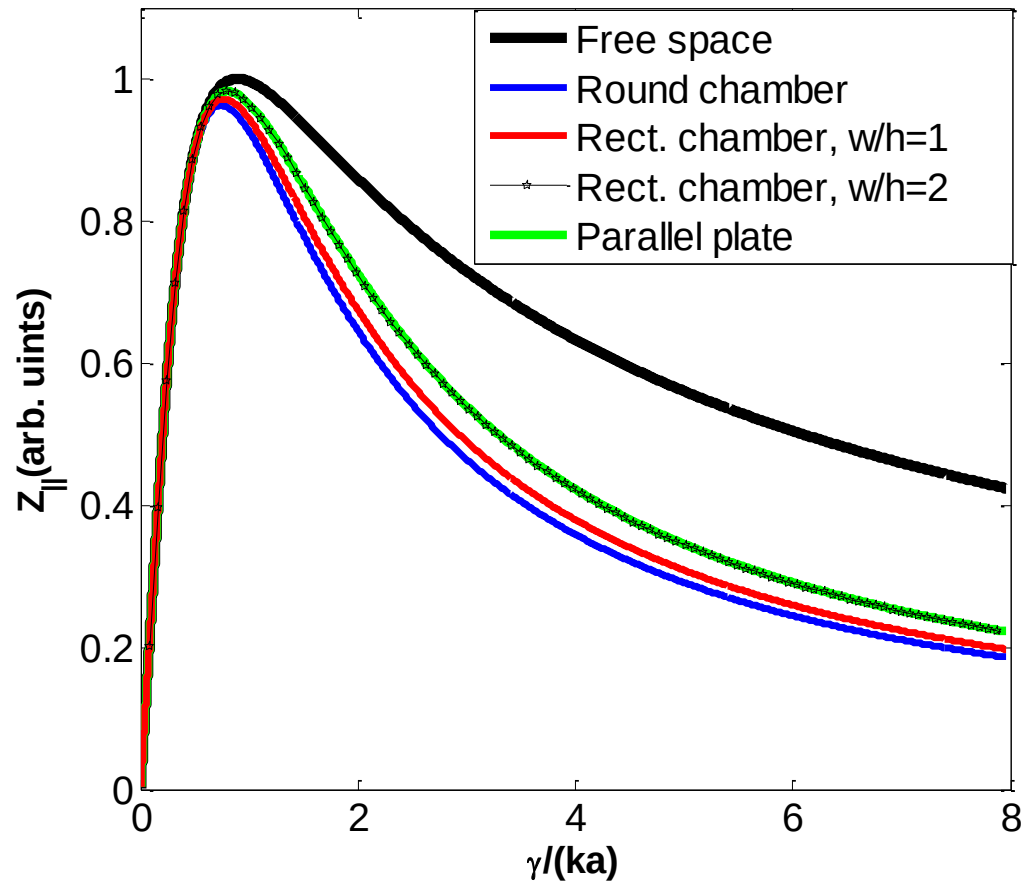

FIG. 12. Comparisons of the shielding effects on the *on-axis* impedance between a round chamber ($r_w = 2a$), parallel plates ($h= 2r_w = 4a$) and a rectangular chamber ($h= 2r_w$) with the ratio $w/h$=1, 2. The full height of the rectangular chamber $h$ is fixed and equal to the diameter of the round chamber.

Our method can also give the LSC impedances averaged over the beam cross-section. Fig. 13 shows the *average* LSC impedance of the round beam with four different boundary conditions: parallel plates, square chamber, round chamber [Eq. (22)], and in free space [Eq. (21)], respectively. The same beam parameters and normalization method are used as in Fig. 12, where the normalized peak impedance in free space is set to 1. The shielding effect on the *average* LSC impedances is similar to the case of *on-axis* LSC impedances. The *average* LSC impedance of the round beam is about 20% less than the *on-axis* one when $1<\gamma/(ka)<8.0$. However, they are almost identical for high frequency with $\gamma/(ka)<1.0$.

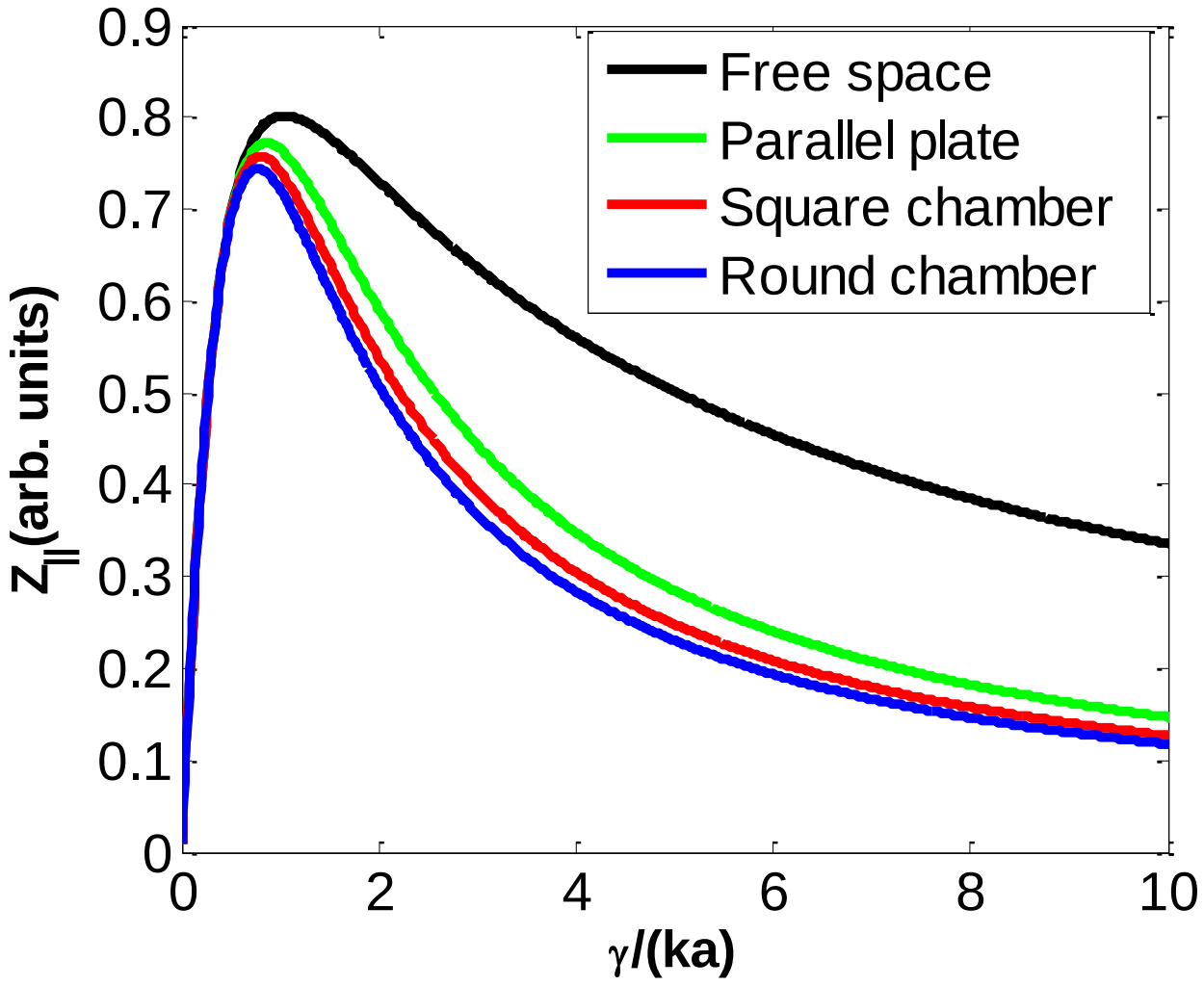

FIG. 13. Comparisons of the shielding effects between the round, square, and parallel-plate chambers on the *average* LSC impedance. The same beam and chamber parameters as in Fig. 12 are used here. The square chamber has an aspect ratio of $w/h$=1.

## VII. CONCLUSIONS

The image charge method is employed in this paper to provide analytical solutions to the full-spectrum LSC impedances of a round uniform beam with line charge modulations inside a rectangular chamber and between parallel plates, respectively. Though the solutions consist of contributions from an infinite number of image beams, due to the fast convergence with the number of image beams, only a finite and small number of image beams are needed to get satisfying resolutions. Since the solutions to the image charge fields induced by each constituent image beam are exact and expressed in a closed form, the resolutions of the total calculated LSC fields and impedances are only dependent on the number of image beams used in the calculation and is insensitive to the ratios of transverse dimensions between the chamber and beam. This is a general method applicable to other beam distributions, such as Gaussian, rectangular and elliptic beams. The calculated LSC impedances in our analysis are valid in the full wavelength spectrum and at arbitrary position within the beam. Moreover, the effect of beam axis offset is also included in our analysis.

Our studies show that the round chambers have slightly better shielding effect than square chambers. When the aspect ratio of a rectangular chamber is larger than 2, the shielding effect is very close to that of a pair of parallel plates. The offset of beam axis slightly reduces the LSC impedance when the offset is small compared to the aperture of beam chamber. However, the reduction becomes significant when the beam is close to the surface of beam chamber.

## ACKNOWLEDGEMENTS

Author Wang would like to thank Prof. Zhirong Huang at SLAC for fruitful discussions.

## REFERENCES

---